\documentclass[11pt,twoside]{article}
\usepackage[utf8]{inputenc}
\usepackage{times,fancyhdr}
\usepackage[dvips]{graphicx}
\usepackage{latexsym}
\usepackage[affil-it]{authblk}
\usepackage{amsmath}
\usepackage{amssymb}
\usepackage{mathtools}
\usepackage[title, titletoc]{appendix}
\newcommand{\KB}{K_{\rm{B}}}
\newcommand{\Ktwo}{\mathcal{K}_2}
\newcommand{\baralpha}{\bar{\alpha}}
\newcommand{\coup}{g_m}
\newcommand{\X}{X}
\newcommand{\Rgal}{R_p}
\newcommand{\dotRgal}{\dot{R}_p}

\usepackage[dvipsnames]{xcolor}
\usepackage[colorlinks,allcolors=Blue]{hyperref}

\usepackage[backend=biber,style=phys,biblabel=brackets,pageranges=false,eprint=true]{biblatex}
\DeclareSourcemap{
 \maps[datatype=bibtex,overwrite=true]{
  \map{
    \step[fieldsource=Collaboration, final=true]
    \step[fieldset=usera, origfieldval, final=true]
  }
 }
}

\renewbibmacro*{author}{%
  \printnames{author}
  \iffieldundef{usera}{
  }{
    (\printfield{usera} Collaboration)
  }
}

\DeclareFieldFormat{eprint:arxiv}{%
    arXiv\addcolon
    \ifhyperref
    {\href{https://arxiv.org/abs/#1}{%
        #1%
    }}
    {%
        #1%
    }}

\AtEveryBibitem{\clearfield{eid}}

\usepackage[top=4cm, bottom=4cm, left=4cm, right=4cm]{geometry}

\begin{document}

\begin{titlepage}

\title{Cherenkov radiation from stars constrains hybrid MOND dark matter models}
\author{Tobias Mistele\thanks{\href{mailto:mistele@fias.uni-frankfurt.de}{mistele@fias.uni-frankfurt.de}}}
\affil{\small Frankfurt Institute for Advanced Studies\\
Ruth-Moufang-Str. 1,
D-60438 Frankfurt am Main, Germany
}
\date{}
\maketitle

\begin{abstract}
We propose a new method to constrain alternative models for dark matter with observations.
Specifically, we consider hybrid models
    in which cold dark matter ({\sc CDM}) phenomena on cosmological scales
    and Modified Newtonian Dynamics ({\sc MOND}) phenomena on galactic scales share a common origin.
Various such models were recently proposed.
They typically contain a mode that is directly coupled to matter (for {\sc MOND})
    and has a non-relativistic sound speed (for {\sc CDM}).
This allows even non-relativistic objects like stars to lose energy through Cherenkov radiation.
This is unusual.
Most modified gravity models have a relativistic sound speed,
    so that only high-energy cosmic rays emit Cherenkov radiation.
We discuss the consequences of this Cherenkov radiation from stars.
\end{abstract}

\end{titlepage}

\section{Introduction}
\label{sec:introduction}

A collisionless fluid is the simplest explanation for the missing mass problem on cosmological scales.
On galactic scales, the simplest explanation is in terms of Modified Newtonian Dynamics ({\sc MOND}) \cite{Milgrom1983a,Milgrom1983b,Milgrom1983c,Bekenstein1984}, i.e. a modified force law.
A natural idea is to find a common origin for both a collisionless fluid on cosmological scales and a MOND-like force on galactic scales in a single model.
Various such models have been proposed.
For example, the superfluid dark matter ({\sc SFDM}) model \cite{Berezhiani2015, Berezhiani2018} and the model by Skordis and Z\l o\'{s}nik ({\sc SZ}) \cite{Skordis2020, Skordis2021}.
We refer to such models as hybrid models.
These typically
    contain a component that plays a role in reproducing both a pressureless fluid on cosmology (CDM) and a modified force in galaxies (MOND).
For example, in {\sc SFDM}, the collisionless fluid on cosmological scales condenses to a superfluid around galaxies.
The phonons of this superfluid then carry a {\sc MOND}-like force.
For our purposes, the important point is that these phonons constitute a massless mode
    which is directly coupled to matter.

\begin{figure}
 \centering
 \includegraphics[width=.3\textwidth]{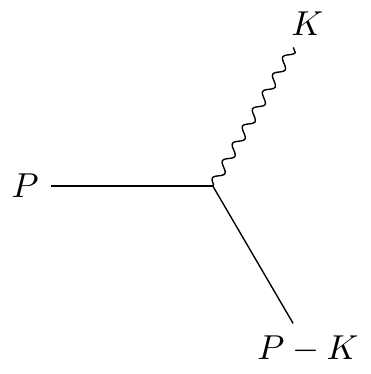}
 \caption{
        Feynman diagram for Cherenkov radiation.
        Straight lines denote a matter object coupled to a massless mode, denoted by a wiggly line.
        The matter object with initial four-momentum $P$ radiates away energy and momentum $K$.
        This process is kinematically allowed when the matter object moves faster than the massless mode propagates.
    }
 \label{fig:cherenkov-feynman}
\end{figure}

Whenever a massless mode is directly coupled to matter, Cherenkov radiation, i.e. the process shown in Fig.~\ref{fig:cherenkov-feynman}, may be possible.
This process is forbidden for slowly-moving matter objects.
But it is allowed for matter objects moving faster than the propagation speed $c_s$ of the massless mode.
In most modified gravity models,
    only relativistic objects emit Cherenkov radiation
    since $c_s$ is relativistic
    \cite{Moore2001, Elliott2005, Bruneton2007b, Milgrom2011, Chesler2017}.
This is different in many hybrid MOND dark matter models since the propagation speed $c_s$ is often non-relativistic.
So even non-relativistic objects like stars can emit Cherenkov radiation.

The reason why $c_s$ is often non-relativistic is as follows.
Any hybrid MOND dark matter model must produce a pressureless fluid on cosmological scales.
Whatever provides this pressureless, i.e. non-relativistic, fluid is often connected to the MOND-like force in galaxies.
Thus, it is natural that the MOND-like force corresponds to a massless mode that propagates with a non-relativistic speed.
For example, as mentioned above, in SFDM the cosmological pressureless fluid condenses to a non-relativistic superfluid around galaxies.
The phonons of this superfluid, which provide the MOND-like force, then have a non-relativistic sound speed $c_s$.
This allows for Cherenkov radiation from non-relativistic objects like stars.
Here, we show how this Cherenkov radiation from stars constrains hybrid {\sc MOND} dark matter models and how such constraints can be avoided.

In Sec.~\ref{sec:general}, we first demonstrate the general idea for a toy model Lagrangian.
Then, we apply our results to standard SFDM in Sec.~\ref{sec:standardsfdm}, to the improved two-field SFDM model from Ref.~\cite{Mistele2020} in Sec.~\ref{sec:twofieldsfdm}, and to the SZ model in Sec.~\ref{sec:relmond}.
We conclude in Sec.~\ref{sec:conclusion}.
In the following, we employ units with $ c = \hbar = 1 $ and the metric signature $ (+, -, -, -) $, unless otherwise stated.
Small Greek indices run from $ 0 $ to $ 3 $ and denote spacetime dimensions.

\section{Toy model}
\label{sec:general}

Consider a real scalar field $\varphi$ that carries a {\sc MOND} force in galaxies, $\varphi \propto \sqrt{G M_{\mathrm{gal}} a_0} \ln(R)$, where $M_{\mathrm{gal}}$ is the mass of the galaxy and $a_0$ is the {\sc MOND} acceleration scale \cite{Bekenstein1984}.
For one of the simplest Lagrangians that produces this behavior, perturbations $\delta$ around such a static galactic background field $\varphi_0$ then have a Lagrangian \cite{Bekenstein1984}
\begin{align}
 \label{eq:prototype}
 \mathcal{L} = \frac12 \frac{1}{\bar{c}^2} (\partial_t \delta)^2 - \frac12 \left( (\vec{\nabla} \delta)^2 + (\hat{a} \vec{\nabla} \delta)^2 \right) - \frac{\coup}{\sqrt{2} M_{\rm{Pl}}} \delta \, \delta_b \,.
\end{align}
Here,
    $\coup$ and $\bar{c}$ are constants that depend on $\varphi_0$,
    and $\delta_b$ is a perturbation of the baryonic density $\rho_b$.
The unit vector $\hat{a}$ points into the direction of the background {\sc MOND} force.
The notation $(\hat{a} \vec{\nabla} \delta)^2$ means $(\sum_{i=1}^3 \hat{a}_i \, \partial_i \delta)^2$.
This form of the Lagrangian is for a uniform background gravitational field, i.e. $\vec{\nabla} \varphi_0 = \mathrm{const}$.
This is a good approximation for perturbations with a sufficiently short wavelength.
Below, we impose cutoffs to ensure that this is indeed the case.
The quantity $\bar{c}$ sets the propagation speed of the perturbation $\delta$.
Indeed, the dispersion relation is $\omega = c_s |\vec{k}|$ with
\begin{align}
  c_s^2 = \bar{c}^2(1 + \gamma^2) \,,
\end{align}
where $\gamma$ is the cosine of the angle between the perturbation's wavevector and $\hat{a}$.

In the original MOND model from Ref.~\cite{Bekenstein1984}, the speed $\bar{c}$ is the speed of light, i.e. the propagation speed is relativistic.
In contrast, as argued above, in hybrid MOND dark matter models, the propagation speed is often non-relativistic
\begin{align}
 \bar{c} \ll 1 \,.
\end{align}
Thus, we take Eq.~\eqref{eq:prototype} with $\bar{c} \ll 1$ as our toy model.
This captures two typical features of galactic-scale perturbations in hybrid models:
There is a direct coupling to matter (for {\sc MOND}) and $c_s$ is non-relativistic (for {\sc CDM}).
As a result, non-relativistic objects like stars emit Cherenkov radiation,
    if their velocity $V$ is larger than a critical velocity $V_{\rm{crit}} = \mathcal{O}(c_s)$.

To make this concrete, consider a star with mass $M$ and velocity $V$ at a distance $\Rgal$ from the center of a galaxy.
For $V > V_{\rm{crit}}$,
    we can calculate the energy loss $\dot{E}_{\rm{Ch}}$ due to Cherenkov radiation in a standard way from the Feynman diagram Fig.~\ref{fig:cherenkov-feynman}, see e.g. Refs.~\cite{Berezhiani2019b, Berezhiani2020}.
The timescale on which stars lose a significant amount of their energy $E$ due to this process is roughly $E/|\dot{E}_{\rm{Ch}}|$.
We define \cite{Mistele2022b}
\begin{align}
 \label{eq:taukin1}
 \tau_E \equiv \frac{E_{\rm{kin}}}{|\dot{E}_{\rm{Ch}}|} \equiv \frac{\frac12 M V^2}{|\dot{E}_{\rm{Ch}}|} = \frac{8 \pi V^3 M_{\rm{Pl}}^2}{f_a \bar{c}^2 \coup^2 M k_{\rm{max}}^2} \, \frac{1}{1 - (k_{\rm{min}}/k_{\rm{max}})^2} \,.
\end{align}
The cutoffs $k_{\rm{max}}$ and $k_{\rm{min}}$ ensure that the Lagrangian Eq.~\eqref{eq:prototype} is valid,
    and $f_a$ depends on the direction of $\hat{a}$ relative to $\vec{V}$.
For circular orbits, $\vec{V} \perp \hat{a}$,
\begin{align}
 f_a = f^\perp_a \equiv \frac{1}{\sqrt{1 + (\bar{c}/V)^2}}\,, \quad
 V_{\mathrm{crit}} = V^\perp_{\mathrm{crit}} \equiv \bar{c}\,.
\end{align}

Stars have both kinetic and potential energy.
For simplicity, $\tau_E$ includes only the kinetic energy.
Still, $\tau_E$ is a useful quantity.
Indeed, for a star with $V > V_{\rm{crit}}$, we have
\begin{align}
 \partial_t \left(E_{\rm{kin}} + E_{\rm{grav}} \right) = \dot{E}_{\rm{Ch}} \,,
\end{align}
with the gravitational energy $E_{\rm{grav}}$.
This gives
\begin{align}
 \label{eq:dotRgal}
 \frac{\dotRgal}{\Rgal} = - \frac{1}{2 \tau_E} \,,
\end{align}
for approximately circular orbits in the {\sc MOND} regime of a galaxy with a flat rotation curve, i.e. $V^2 = \sqrt{G M_{\rm{gal}} a_0}$ \cite{Mistele2022b}.
Thus,
    if $\tau_E$ depends only weakly on $\Rgal$,
    stars transition to smaller galactic radii as $\exp(-t/2\tau_E)$ due to Cherenkov radiation.

\begin{figure}
 \centering
 \includegraphics[width=.49\textwidth]{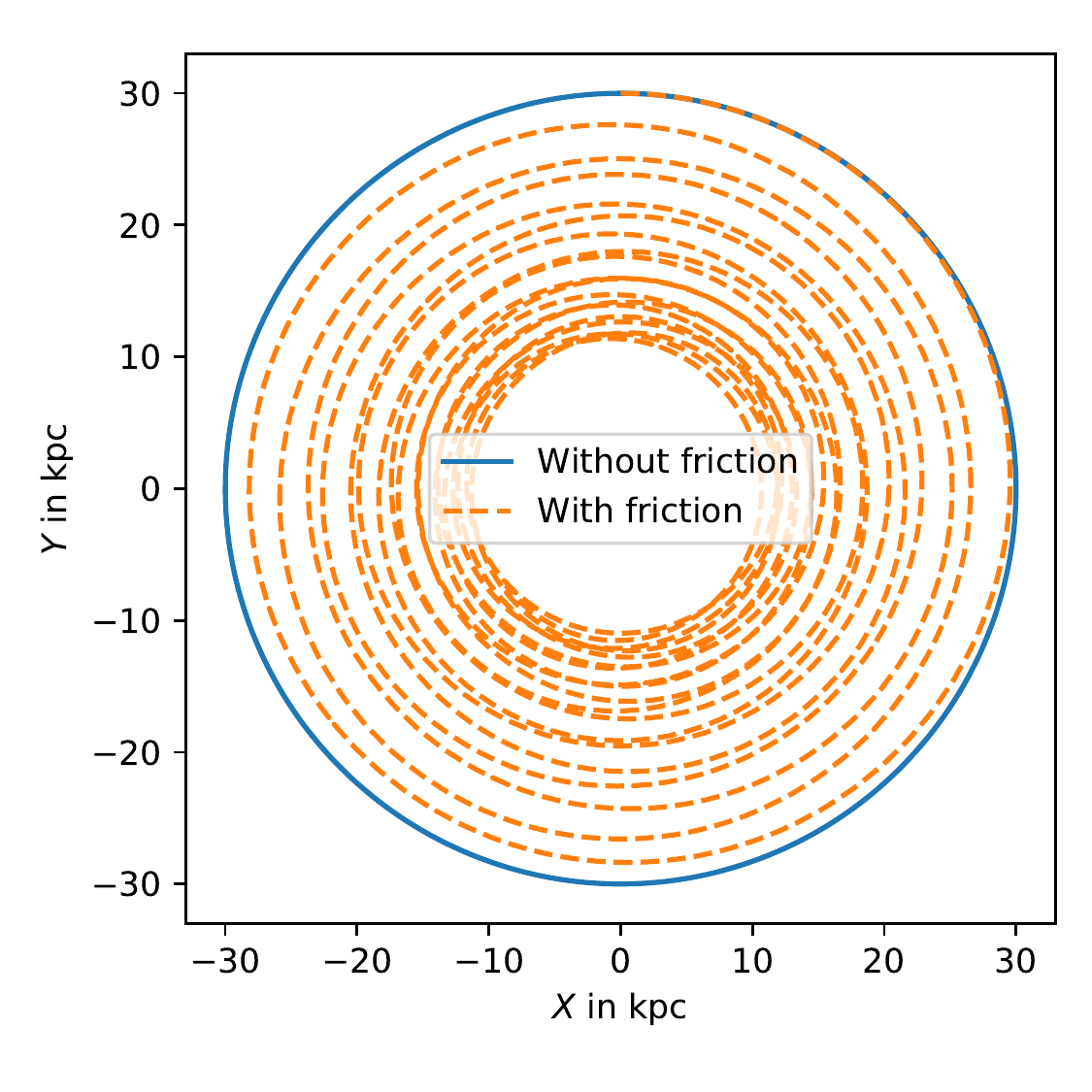}
 \caption{
     Orbits of a star in the MOND regime of a galaxy with initial conditions $R_0 = 30\,\mathrm{kpc}$ and $V_0 = 200\,\rm{km}/\rm{s}$ without (solid blue line) and with (dashed orange line) energy loss due to Cherenkov radiation.
     The energy loss through Cherenkov radiation corresponds to
        $\tau_E/V^3 = 5 \cdot 10^9\,\rm{yr} / (200\,\mathrm{km}/\mathrm{s})^3$.
     The galaxy mass is chosen such that the orbit is circular without Cherenkov radiation.
 }
 \label{fig:orbitXY}
\end{figure}

This is confirmed by a numerical analysis of test particle orbits in a galaxy with a friction force corresponding to the Cherenkov radiation energy loss $\dot{E}_{\mathrm{Ch}}$ \cite{Mistele2022b}.
We show an example of such an orbit in Fig.~\ref{fig:orbitXY}.
The orbital decay is due to a friction force producing an energy loss $\dot{E}_{\rm{Ch}}$ with, initially, $\tau_E = 5 \cdot 10^9\,\rm{yr}$.
The initial conditions are such that the orbit is circular without Cherenkov radiation.
Treating this Cherenkov radiation as a friction force is justified
    because the energy loss happens through a large number of emissions,
    each carrying only a small fraction of the star's energy.
This is due to the strict cutoffs we impose (see below).
In Fig.~\ref{fig:orbitXY}, the friction force acts in the direction of $\vec{V}$.
We have numerically verified that other directions give similar results.
For Fig.~\ref{fig:orbitXY}, we have further assumed that $\tau_E$ is independent of $\Rgal$.
If $\tau_E$ depends on $\Rgal$, the orbital decay is no longer exponential.
But we have numerically verified that the orbital decay still happens on a timescale $\tau_E$ and is accurately captured by Eq.~\eqref{eq:dotRgal}.
We have further verified that other initial conditions give similar results, at least as long as the orbit without the friction force is still close to circular.

We now choose the cutoffs $k_{\rm{min}}$ and $k_{\rm{max}}$.
Since the background galaxy's field varies on $\rm{kpc}$ scales, we choose
\begin{align}
 k_{\rm{min}} \sim 1/\mathrm{kpc} \sim 10^{-26}\,\rm{eV} \,.
\end{align}
For $k_{\rm{max}}$, we choose \cite{Mistele2022b}
\begin{align}
 k_{\rm{max}} &\sim f_p \cdot r_{\mathrm{MOND}}^{-1} \cdot \sqrt{\frac{a_b^{\rm{gal}}}{a_0}} \sim 10^{-22}\,\mathrm{eV} \cdot f_p \cdot \sqrt{\frac{a_b^{\rm{gal}}}{a_0}}  \,,
\end{align}
where $a_b^{\rm{gal}}$ is the Newtonian baryonic acceleration of the galaxy at the star's position,
    $r_{\mathrm{MOND}} = \sqrt{G M/a_0}$ is the star's MOND radius,
    and $f_p$ parametrizes additional model-dependent cutoffs.
For standard {\sc SFDM} and $\vec{V} \perp \hat{a}$, we have explicitly verified that we can take $f_p = 1$ \cite{Mistele2022b}.
For the numerical value on the right-hand side we assumed $M = M_\odot$.
This choice of $k_{\rm{max}}$ avoids two complications close to the star.
First, sufficiently close to the star, the star's own field is no longer small compared to the background galaxy's field so we cannot treat it as a perturbation.
Second, the acceleration due to the star becomes larger than $a_0$.
In this high-acceleration regime, some models postulate different behavior such as higher-derivative terms becoming important \cite{Babichev2011, Berezhiani2015, Skordis2020} which would complicate the calculation.

With our particular choice of $k_{\rm{max}}$ and using $k_{\rm{min}} \ll k_{\rm{max}}$, we find
\begin{align}
 \tau_E = \frac{2\cdot10^8\,\mathrm{yr}}{f_a f_p^2 \, \coup^2} \cdot \left(\frac{V/\bar{c}}{2}\right)^2 \cdot \left(\frac{a_0}{a_b^{\rm{gal}}}\right) \cdot \left(\frac{V}{200\,\rm{km}/\rm{s}}\right) \cdot \left(\frac{1.2 \cdot 10^{-10}\,\rm{m}/\rm{s}^2}{a_0}\right) \,.
\end{align}
Thus, for $\coup$ of order 1 and $V > V_{\rm{crit}}$, stars lose a significant fraction of their energy on roughly galactic timescales.
This is a conservative estimate.
The actual energy loss may be higher, e.g. $k > k_{\rm{max}}$ modes may contribute, but are not considered here.

\section{Application to standard {\sc SFDM}}
\label{sec:standardsfdm}

The {\sc SFDM} model from Ref.~\cite{Berezhiani2015} introduces a new type of particle which behaves like standard cold dark matter on cosmological scales.
On galactic scales, it condenses to a superfluid whose phonons mediate a {\sc MOND}-like force.
The phonon field $\theta$ is responsible for both the superfluid and the {\sc MOND} force.
It is described by an effective Lagrangian
\begin{align}
 \mathcal{L} = \frac{2\Lambda}{3} (2m)^{3/2} \sqrt{|X - \beta Y|} X - \frac{\baralpha \Lambda}{M_{\rm{Pl}}} \rho_b \, \theta \,,
\end{align}
with
\begin{align}
 X = \dot{\theta} + \hat{\mu} -  (\vec{\nabla} \theta)^2/(2m) \,, \quad Y = \dot{\theta} + \hat{\mu} \,, \quad \hat{\mu} = \mu_{\rm{nr}} - m \phi_{\rm{N}} \,.
\end{align}
Here, $m$ is the mass of the particles, $\mu_{\rm{nr}}$ is the non-relativistic chemical potential, $\phi_N$ is the Newtonian gravitational potential, $\beta$ parametrizes finite-temperature effects, and $\Lambda$  and $\baralpha$ are constants.
To avoid an instability and for a positive superfluid energy density, we need $\beta \in (3/2, 3)$ \cite{Berezhiani2015}.

Consider a galaxy in the {\sc MOND} limit, $ (\vec{\nabla} \theta)^2 \gg 2 m \hat{\mu} $ \cite{Berezhiani2015}.
Up to a term that mixes spatial and time derivatives, the Lagrangian for perturbations on top of such a galaxy has the form of Eq.~\eqref{eq:prototype} with
\begin{align}
 \bar{c} =  3 \bar{f}_\beta \, \frac{|\vec{a}_{\theta}|}{a_0} \frac{\sqrt{\baralpha}}{m} \sqrt{a_0 M_{\mathrm{Pl}}}\,, \;
 \coup = \sqrt{\frac{a_0}{|a_{\theta}|}} \,, \;
 \bar{f}_\beta =  \frac{1}{\sqrt{3 (\beta-1)(\beta+3)}} \,,
\end{align}
where $a_0 = \baralpha^3 \Lambda^2/M_{\rm{Pl}}$, $\vec{a}_\theta = - (\baralpha \Lambda/M_{\mathrm{Pl}}) \vec{\nabla} \theta$, and $\hat{a} \propto \vec{\nabla} \theta$ \cite{Mistele2022b, Berezhiani2015}.
The sound speed is typically non-relativistic.
For the fiducial parameters from Ref.~\cite{Berezhiani2018},
$ \bar{c} = 375\,\mathrm{km}/\mathrm{s} \cdot (a_{\theta}/a_0) $.

The term that mixes spatial and time derivatives makes a standard calculation based on the Feynman diagram Fig.~\ref{fig:cherenkov-feynman} more complicated.
So we do a classical calculation instead \cite{Mistele2022b}, following Ref.~\cite{Jackson1998}.\footnote{
   We expect classical and quantum calculations to give the same result since the Feynman diagram Fig.~\ref{fig:cherenkov-feynman} does not contain any loops.
   Higher-order corrections would likely lead to differences.
   But here we are only interested in the leading order effect corresponding to Fig.~\ref{fig:cherenkov-feynman}.
}
The result has the same form as before, but with adjusted $f_a$ and $V_{\rm{crit}}$.
For $\vec{V} \perp \hat{a}$,
        \begin{align}
         V_{\rm{crit}}^\perp = \bar{c} \, \sqrt{\frac{2}{2 + f_\beta^2}} \,, \quad
         f_a^\perp = \frac{1}{\sqrt{2}} \frac{1}{1 + f_\beta^2} \,,
        \end{align}
where $f_\beta = (3 - \beta) \bar{f}_\beta$.
Because of a more conservative approximation to keep the calculation with $f_\beta \neq 0 $ simple,
    $f_a^\perp$ does not reproduce the previous $f_\beta = 0$ result \cite{Mistele2022b}.

\begin{figure}
 \centering
 \includegraphics[width=.7\textwidth]{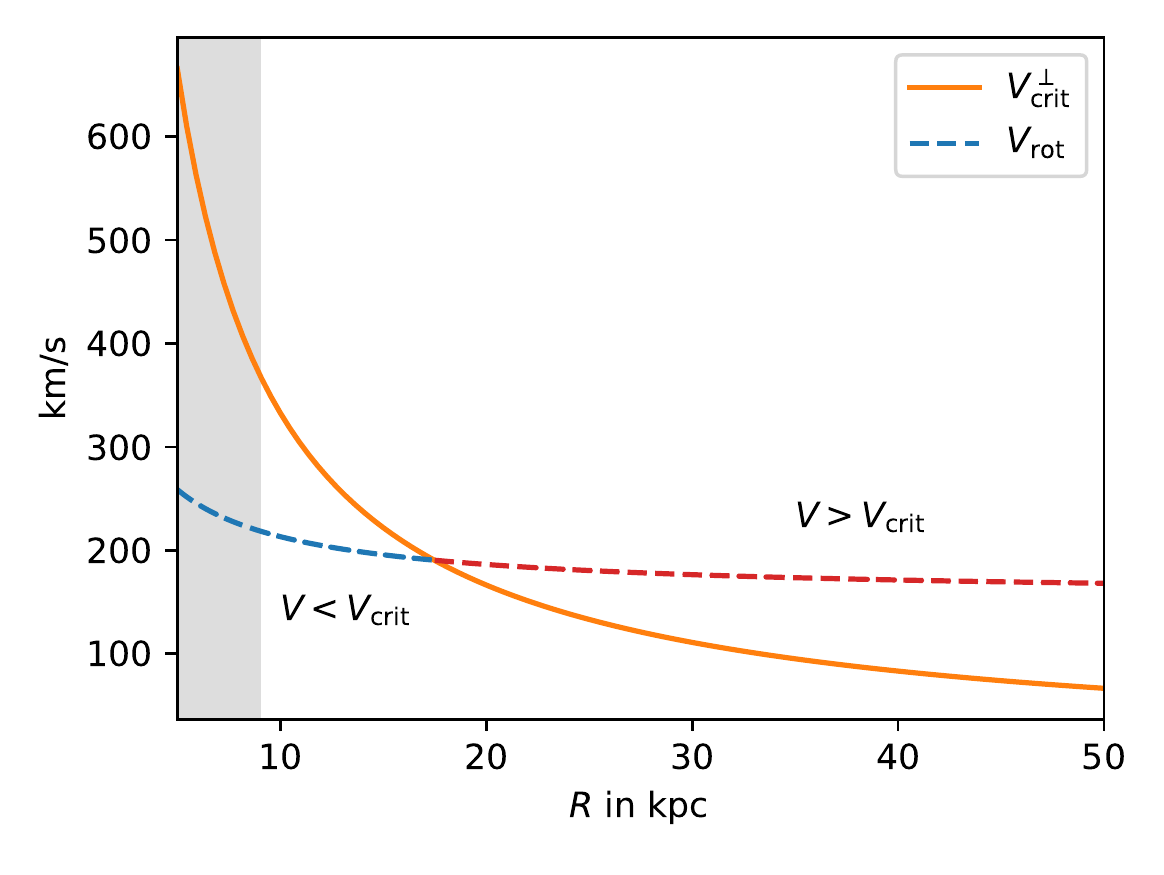}
 \caption{
     Critical velocity and rotation curve of a galaxy with mass $M_{\rm{gal}} = 5 \cdot 10^{10}\,M_\odot$ concentrated at its center in standard {\sc SFDM} with the fiducial parameters from Ref.~\cite{Berezhiani2018}.
     The shaded region is not in the {\sc MOND} regime since $a_b^{\rm{gal}} > a_0$.
    }
 \label{fig:sfdm-vrot-vs-soundspeed}
\end{figure}

In the {\sc MOND} limit, we have $a_\theta \approx \sqrt{a_0 a_b^{\mathrm{gal}}}$.
Thus, we have  $\bar{c} \propto 1/\Rgal$ and $V_{\rm{crit}}^\perp \propto 1/\Rgal$.
Since rotation curves are flat at large radii,
    there is a critical radius where $V_{\rm{crit}}$ drops below the rotation curve $V_{\rm{rot}}$.
Beyond this radius,
    stars with velocity $V_{\rm{rot}}$ lose energy on timescales $\tau_E$,
    see Fig.~\ref{fig:sfdm-vrot-vs-soundspeed}.

We can then use the Milky Way stellar rotation curve to rule out the MOND limit of standard {\sc SFDM} in the Milky Way,
    at least for a range of parameter values.
Concretely,
    the Milky Way cannot be in the MOND limit of standard {\sc SFDM}
    unless either $\bar{c}$ is large enough to kinematically forbid Cherenkov radiation, i.e. $V_{\rm{rot}} < V_{\rm{crit}}^\perp$,
    or $\tau_E$ is larger than galactic timescales, i.e. $\tau_E > \tau_{\rm{min}}$ for some $\tau_{\rm{min}}$.
Since $\tau_E \propto 1/\bar{c}^2$, the latter can be achieved by making $\bar{c}$ small.
Here, we assume $M_{\rm{gal}} = 6 \cdot 10^{10}\,M_\odot$ \cite{Mistele2020, Hossenfelder2020}
    and the rotation curve from Refs.~\cite{McGaugh2019b, Eilers2019}.
We choose $\tau_{\rm{min}} = 10^{10}\,\rm{yr}$, i.e. stars should not lose much energy in $10^{10}\,\rm{yr}$.
Then, for a given $\beta$, the rotation curve at each point $\Rgal$ excludes an interval of $\sqrt{\baralpha}/m$.
Concretely,
    the rotation curve at a radius $\Rgal$ rules out
    $\sqrt{\baralpha}/m$ in the interval
        \begin{align}
         \label{eq:standardsfdm:rotcurve:interval1}
         \frac{V_{\mathrm{rot}} \Rgal}{\bar{f}_\beta} \frac{\sqrt{8 \pi}}{3} \sqrt{\frac{M_{\mathrm{Pl}}}{M_{\mathrm{gal}}}} \cdot
         \left(
             \sqrt{\frac{V_{\mathrm{rot}} R_p \sqrt{2} ( 1 + f_\beta^2 )}{\tau_{\mathrm{min}} \sqrt{a_0 G M_{\mathrm{gal}}}}},
             \sqrt{1 + \frac12 f_\beta^2}
          \right) \,,
        \end{align}
    where $a \cdot (x_1, x_2) \equiv (a x_1, a x_2)$.
At the lower boundary of this interval,
    we will use the fixed but unusually low value $a_0 = 0.5 \cdot 10^{-10}\,\mathrm{m}/\mathrm{s}^2$ instead of $a_0 = \baralpha^3 \Lambda^2/M_{\rm{Pl}}$
    to set a conservative limit and keep things simple.

For $\beta \in \{3/2, 2, 3\}$, we list these intervals from different radii of the Milky Way rotation curve in Table~\ref{tab:standardsfdm:constraints}.
Together, they rule out $\sqrt{\bar{\alpha}}/m$ in an interval \cite{Mistele2022b}
    \begin{align}
     \label{eq:standardsfdm:rotcurve:intervalnumeric}
     \sqrt{\frac{6\cdot10^{10}\,M_\odot}{M_{\rm{gal}}}} \left(
         \sqrt{\frac{10^{10}\,\rm{yr}}{\tau_{\rm{min}}}} \left(\frac{6\cdot 10^{10}\,M_\odot}{M_{\rm{gal}}}\right)^{1/4} q_l,
         q_h
         \right) \cdot \rm{eV}^{-1} \,,
    \end{align}
    for some $q_l$ and $q_h$.
Concretely,
    \begin{align}
    \label{eq:standardsfdm:q}
    \begin{split}
     q_l &= 0.25\,, q_h = 2.34\,, \quad \mathrm{for\,} \beta =3/2 \,, \\
     q_l &= 0.34\,, q_h = 3.29\,, \quad \mathrm{for\,} \beta =2 \,, \\
     q_l &= 0.51\,, q_h = 5.01\,, \quad \mathrm{for\,} \beta =3 \,.
    \end{split}
    \end{align}
This also excludes the fiducial value $\sqrt{\baralpha}/m \approx 2.4\,\rm{eV}^{-1}$ for $\beta=2$ from Ref.~\cite{Berezhiani2018}.

Of course, this only excludes the MOND limit $(\vec{\nabla} \theta)^2 \ll 2 m \hat{\mu}$ of {\sc SFDM}, not the model in general.
However, this MOND limit is one of the main motivations behind {\sc SFDM}.
It is important to know when this limit can and cannot exist.

Another caveat is that this only excludes the MOND limit in the Milky Way.
In principle, it could be that most other galaxies \textit{can} be in the MOND limit of {\sc SFDM}.
We don't expect this to be the case since the Milky Way is not that special.
Still, in future work, the above analysis should be repeated with a larger sample of galaxies with resolved stellar rotation curves.

\begin{table}
 \centering
 \begin{tabular}{c|c|c|c|c}
  $\Rgal$ &
  $V$ &
  $(q_l, q_h)$ for $\beta=3/2$ &
  $(q_l, q_h)$ for $\beta=2$ &
  $(q_l, q_h)$ for $\beta=3$ \\
  $\rm{kpc}$ &
  $\rm{km}/\rm{s}$ &
  &
  &
  \\
  \hline
  ${ 15.2 }$ & ${ 220 }^{ +1 }_{ -1 }$ & $\left( 0.25, 1.56 \right)$ & $\left( 0.34, 2.19 \right)$ & $\left( 0.51, 3.34 \right)$ \\
${ 20.3 }$ & ${ 203 }^{ +3 }_{ -3 }$ & $\left( 0.35, 1.92 \right)$ & $\left( 0.46, 2.70 \right)$ & $\left( 0.69, 4.11 \right)$ \\
${ 24.8 }$ & ${ 202 }^{ +6 }_{ -6 }$ & $\left( 0.47, 2.34 \right)$ & $\left( 0.62, 3.29 \right)$ & $\left( 0.93, 5.01 \right)$
 \end{tabular}
 \caption{
     Excluded $\sqrt{\bar{\alpha}}/m$ intervals $(q_l, q_h) \cdot \rm{eV}^{-1}$ from the Milky Way rotation curve at different radii.
 }
 \label{tab:standardsfdm:constraints}
\end{table}

\section{Application to two-field {\sc SFDM}}
\label{sec:twofieldsfdm}

To avoid tensions within standard {\sc SFDM}, Ref.~\cite{Mistele2020} proposed a model
    with phenomenology close to the original {\sc SFDM} model,
    but in which the two roles of the field $\theta$
    are split between two different fields.
The field $\phi_- = \rho_- e^{-i \theta_-}/\sqrt{2}$ carries the superfluid's energy density,
    but is not directly coupled to normal matter.
In contrast,
    the field $\theta_+$ is coupled directly to normal matter and carries a {\sc MOND}-like force in equilibrium.
The Lagrangian reads
\begin{align}
\mathcal{L} = \mathcal{L}_- + f(K_+ + K_- - m^2) - \frac{\baralpha \Lambda}{M_{\rm{Pl}}} \, \theta_+\, \rho_b \,,
\end{align}
where $K_\pm = \nabla_\alpha \theta_\pm \nabla^\alpha \theta_\pm$ and
\begin{align}
\mathcal{L}_- = (\nabla_\alpha \phi)^*(\nabla^\alpha \phi) - m^2 |\phi|^2 - \lambda_4 |\phi|^4 \,.
\end{align}
The function $f(K) \equiv \sqrt{|K|} K$ is similar to standard SFDM but contains both $\theta_+$ and $\theta_-$.

There are two massless modes, roughly corresponding to $\theta_+$ and $\theta_-$.
Only the $\theta_-$ mode is relevant for us because it has a non-relativistic sound speed,
\begin{align}
 c_s = \sqrt{\frac{\hat{\mu}}{m}} \ll 1\,.
\end{align}
This mode couples to normal matter only indirectly through a mixing of $\theta_-$ and $\theta_+$ from the $f(K_+ + K_- - m^2)$ term.
This suppresses Cherenkov radiation, \cite{Mistele2022b}
\begin{align}
 g_m = \mathcal{O}\left(\frac{\sqrt{\lambda_4}}{\baralpha}\right) \ll 1\,,
\end{align}
so that
\begin{align}
 \tau_E
    \sim \frac{\baralpha^2}{\lambda_4} \left(\frac{V}{c_s}\right)^2  \frac{V}{a_0}
    \sim 10^{16}\,\mathrm{yr} \left(\frac{V}{200\,\rm{km}/\rm{s}}\right) \left(\frac{V}{c_s}\right)^2 \sqrt{\frac{10^{-2} a_0}{\bar{a}}} \,.
\end{align}
Here, $\bar{a} \ll a_0$ is an acceleration below which the equilibrium becomes unstable \cite{Mistele2020}.
Thus, $\tau_E$ is much larger than the age of the universe and does not constrain the model.
The reason is that the non-relativistic massless mode couples to normal matter only indirectly through a mixing.

\section{Application to the {\sc SZ} model}
\label{sec:relmond}

Skordis and Z\l o\'{s}nik have recently proposed a hybrid {\sc MOND} dark matter model based on a scalar field $\phi$ and a unit vector field $A_\mu$ \cite{Skordis2020, Skordis2021}.
On cosmological scales, the scalar field $\phi(t)$ is involved in providing a {\sc CDM}-like fluid.
In galaxies, $\phi = Q_0 \cdot t + \varphi$ where $Q_0$ is constant and $\varphi$ carries the {\sc MOND}-like force.
Thus, $\phi$ plays a double role analogous to the phonon field in SFDM and, potentially, stars can emit Cherenkov radiation.

For simplicity, like Refs.~\cite{Skordis2020, Skordis2021}, we consider perturbations on top of the late-time Minkowski limit $\phi = Q_0 t$ and not on top of a galaxy.
We assume that our results are qualitatively valid also in galaxies.
For Cherenkov radiation, we are interested in dynamic propagating modes.
There is one scalar mode involving $\phi$ that has a potentially non-relativistic sound speed \cite{Skordis2020, Skordis2021}
\begin{align}
 c_s = \sqrt{\frac{\left(2 - \KB\right) \left(1 + \frac12 \lambda_s \KB\right)}{\Ktwo \KB}} \,,
\end{align}
where $\Ktwo$, $\lambda_s$, and $\KB$ are parameters of the model.
Strictly speaking, this mode has a dispersion relation $\omega^2 = c_s^2 k^2 + \mathcal{M}^2$ and so is a massive mode.
However, for the wavevectors we consider here, $k \gtrsim 1/\mathrm{kpc}$, the mass term is negligible.
This is because the condition $\mathcal{M} \ll c_s k$ is equivalent to
\begin{align}
 \frac{\mathcal{M}^2}{c_s^2 k^2} = \frac{m_{\mathrm{SZ}}^2}{k^2} \frac{2-\KB}{2} \frac{1 + \lambda_s}{1 + \frac12 \lambda_s \KB} \ll 1 \,,
\end{align}
which is always fulfilled in our case.
To see this, first note that $0 < \KB < 2$ is required for stability \cite{Skordis2020} and $\lambda_s$ is small in the MOND limit in which we are interested here \cite{Skordis2020}.
Thus, the size of $\mathcal{M}/c_s k$ is mainly determined by the ratio $m_{\mathrm{SZ}}/k$ with the mass parameter $m_{\mathrm{SZ}} = \sqrt{2 \Ktwo/(2-\KB)} Q_0$.
This ratio $m_{\mathrm{SZ}}/k$ is small because a MOND-like force on galactic scales requires the mass parameter $m_{\mathrm{SZ}}$ to be smaller than about  $1/\mathrm{Mpc}$ \cite{Skordis2020}.

Thus, Cherenkov radiation from stars seems to be possible in this model,
    at least for $c_s \ll 1$.
However, it turns out that Cherenkov radiation is actually strongly suppressed.
The reason is that the coupling of the scalar mode to matter vanishes in dynamical situations when evaluated on-shell,
    i.e. when evaluated for $\omega^2 = c_s^2 k^2 + \mathcal{M}^2$.
This is in contrast to the static limit where $\varphi$ must have a standard gravitational coupling to matter in order to mediate a MOND-like force.
We will now explain this in a bit more detail.

The Lagrangian of the SZ model contains terms \cite{Skordis2020, Skordis2021}
\begin{align}
 \mathcal{L} = \KB \left( \dot{\vec{A}} + \vec{\nabla} \phi_N \right)^2 + 2 (2 - \KB) \left(\dot{\vec{A}} + \vec{\nabla} \phi_N \right) \cdot \vec{\nabla} \varphi + \dots \,.
\end{align}
To see where the matter coupling of $\varphi$ comes from, consider the $\varphi$ equation of motion in the static limit.
Roughly, we have
\begin{align}
 0 = \dots + \vec{\nabla}^2 \phi_N \,.
\end{align}
In addition, the $\phi_N$ equation of motion gives $\vec{\nabla}^2 \phi_N \propto \rho_b/M_{\rm{Pl}}^2 + \dots$
which introduces a source term $\rho_b/M_{\mathrm{Pl}}^2$ in the $\varphi$ equation of motion.
This is how $\varphi$ is coupled to normal matter in the static limit which allows it to mediate a MOND-like force.

Consider now again the $\varphi$ equation but without setting time derivatives to zero,
\begin{align}
 0 = \dots + \vec{\nabla} \left( \dot{\vec{A}} + \vec{\nabla} \phi_N \right) \,.
\end{align}
The combination $\dot{\vec{A}} + \vec{\nabla} \phi_N$ also occurs in the $\vec{A}$ equation of motion,
\begin{align}
 0 = \dots + \partial_t \left(\dot{\vec{A}} + \vec{\nabla} \phi_N \right) \,,
\end{align}
where, for scalar perturbations, all $\nabla^2 A$ terms cancel \cite{Skordis2020}.
Then, the $\ddot{\vec{A}}$ term may dominate in this equation although, for a nonrelativistic sound speed, the dispersion relation $\omega \approx c_s k \ll k$ implies that time derivatives are much smaller than spatial derivatives.
Thus, we have $\dot{\vec{A}} \approx - \vec{\nabla} \phi_N + \dots$,
    which cancels the $\vec{\nabla}^2 \phi_N$ term,
    and therefore the matter coupling,
    in the $\varphi$ equation of motion.
A more careful calculation shows that the coupling to matter vanishes when evaluated on-shell, i.e. when evaluated for $\omega^2 = c_s^2 k^2 + \mathcal{M}^2$ \cite{Mistele2022b}.

Since the matter coupling is evaluated on-shell in the leading-order Feynman diagram (see Fig.~\ref{fig:cherenkov-feynman}), the leading order Cherenkov radiation vanishes in this model.
We do not expect higher-order corrections to be significant.
As a result, this model is not constrained by Cherenkov radiation from stars.
The reason is the suppressed matter coupling in dynamical situations.

We expect that this suppressed matter coupling is also relevant beyond Cherenkov radiation constraints.
For example regarding energy loss constraints from binary pulsars.
But note that it might not always help in matching observations.
For example, most successful predictions of MOND assume an instantaneous force.
Naively, this is justified on timescales larger than $d/c_s$ where $d$ is the size of the spatial region under consideration.
If $c_s$ is sufficiently small, this is a concern for hybrid models in general.
But in the SZ model in particular, the MOND force might take even longer to reach its \mbox{(quasi-)static} limit due to the suppressed matter coupling of the scalar mode.
That is, assuming an instantaneous force might be valid only on even longer timescales.
Thus, even if $c_s$ is sufficiently large, the SZ model might not reproduce the successes of MOND.

However, one should be careful with these heuristics.
For example, as discussed above, the MOND force in the SZ model involves a mixing of the metric $g_{\mu \nu}$ and the scalar field $\varphi$.
Thus, in principle, the tensor mode, whose coupling is not suppressed, might be more relevant than the scalar mode for reaching the (quasi-)static limit.
That is, it's not clear how much the suppressed matter coupling of the scalar mode matters.
This requires further investigation that we leave for future work.

\section{Conclusion}
\label{sec:conclusion}

One usually avoids superluminal sound speeds for theoretical reasons.
Yet, for empirical reasons, subluminal sound speeds are also dangerous,
    since these allow for Cherenkov radiation.
Hybrid {\sc MOND} dark matter models with a common origin for cosmological and galactic phenomena often allow Cherenkov radiation even for non-relativistic objects like stars.
We have shown how this rules out some of the parameter space of standard {\sc SFDM}
    despite restrictive cuts to avoid technical complications.
We also discussed how one may evade these constraints, namely by mixing (two-field {\sc SFDM}) or a suppressed matter coupling in dynamical situations ({\sc SZ} model).

\section*{Acknowledgements}
\label{sec:acknowledgements}

I am grateful for financial support from FIAS.
I thank Sabine Hossenfelder, Stacy McGaugh, and Luciano Rezzolla for discussions.

\printbibliography[heading=bibintoc]

\end{document}